\documentclass[sigconf,natbib=true,anonymous=false]{acmart}
\usepackage{algorithm}
\usepackage{algorithmic}
\usepackage{amsmath}
\usepackage{natbib}      
\usepackage{threeparttable}
\usepackage{makecell}
\AtBeginDocument{%
  }

\acmConference[CIKM]{International Conference on Information and Knowledge Management}{Nov 10-14,
  2025}{SEOUL, KOREA}

\begin{document}

\title{Reinforcing User Interest Evolution in Multi-Scenario Learning for recommender systems}


\author{Zhijian Feng}
\email{zhijian.feng.ai@gmail.com}
\affiliation{%
  \institution{Unaffiliated}
  \city{}
  \country{China}
}


\author{Wenhao	Zheng}
\email{zhijian.feng.ai@gmail.com}
 \affiliation{
   \institution{Microsoft Software Technology Center Asia}
   \city{}
   \country{China}
 }

\author{Xuanji	Xiao}
\email{xuanji.xiao@gmail.com}
 \affiliation{%
  \institution{Unaffiliated}
  \city{}
  \country{China}}





\renewcommand{\shortauthors}{}
\begin{abstract}
In real-world recommendation systems, users would engage in variety scenarios, such as homepages, search pages, and related recommendation pages. Each of these scenarios would reflect different aspects users focus on. However, the user interests may be inconsistent in different  scenarios, due to differences in decision-making processes and preference expression. This variability complicates unified modeling, making multi-scenario learning a significant challenge. To address this, we propose a novel reinforcement learning approach that models user preferences across scenarios by modeling user interest evolution across multiple scenarios. Our method employs Double Q-learning to enhance next-item prediction accuracy and optimizes contrastive learning loss using Q-value to make model performance better. Experimental results demonstrate that our approach surpasses state-of-the-art methods in multi-scenario recommendation tasks. Our work offers a fresh perspective on multi-scenario modeling and highlights promising directions for future research.
\end{abstract}

\maketitle

\section{Introduction}

With the continuous exponential growth of data, DLRM (Deep Learning Recommendation Models) are gradually becoming one of the core parts for major Internet platforms, which could enhance user experience and business value. In the design and optimization of DLRM, improving user engagement has become the key objective. 
To achieve this goal, an accurate identification of user interests is crucial. Through this way, the recommendation system could deliver the content that meets users need.

Traditional methods have made some progress in identifying user interests.
Wide \& Deep\citep{cheng2016wide} combines linear and deep models to balance memorization and generalization to enhance the modeling of user interest. 
DeepFM\citep{guo2017deepfm} and DCN\citep{wang2017deep} extends this by modeling user interest interactions with incorporating factorization machines and cross networks respectively. DIN\citep{zhou2018deep} introduces attention mechanisms to adaptively weight user interests for fine-grained personalization. Above methods improve the accuracy and interpretability of user interest identification.


\begin{figure}[t]
\centering
\includegraphics[width=0.8\linewidth]{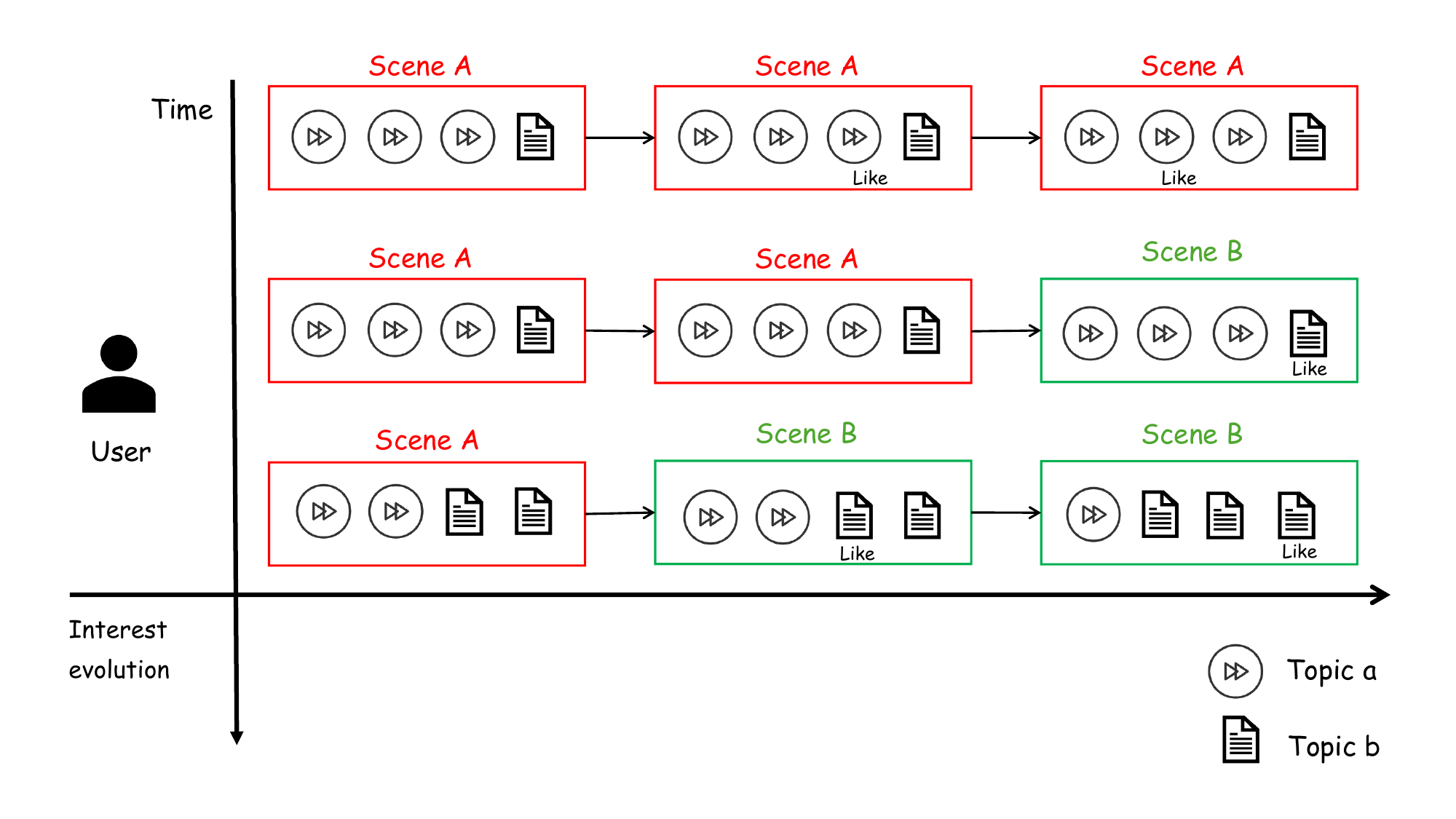}
\caption{The user starts browsing in Scene A with an interest in topic a, but over time, the user's interest shifts to Scene B, and his interest also shifts to topic b. At this point, Scene A would increase the display of topic b, but since the user's focus is no longer on Scene A, the results shown in Scene A are considered as exposure bias, which would be an challenge in multi-scenario learning.}
\label{fig:example}
\vspace{-10pt}
\end{figure}

However, real-world multiple scenarios would show different contents for the same user to satisfy user diverse interest, which could make user spend more time in platform. It would lead to disperse user interests, making it challenging to simultaneously improve performance across all scenarios.
Hence, multi-scenario modeling in DLRM is gradually becoming a research hotspot. For example, STAR\citep{sheng2021one} proposes a multi-scenario recommendation framework based on multi-task learning, which optimizes multiple recommendation tasks simultaneously by sharing underlying feature representations.
PEPNet\citep{chang2023pepnet} proposes a multi-scenario recommendation method based on hierarchical information extraction, which achieves cross-scenario knowledge transfer and collaborative optimization through hierarchical feature extraction and task-sharing mechanisms. These studies show that multi-scenario modeling alleviates data sparsity and captures user preference shifts, enabling richer semantics and more accurate recommendations.

The previous works fail to address two key issues. First, user interests could shift across different scenarios. Second, a user's interests in different scenarios may exhibit negative transfer. As shown in Fig~\ref{fig:example}, user interests gradually shift from Scene A to Scene B over time. In Scene A, users preferred Topic a. However, as time progressed, their interests gradually shifted toward Topic b in Scene B. Despite this transition, Scene B failed to promptly detect the shift in user interest. Meanwhile, Scene A began to exhibit unexpected interest patterns, likely due to interference from Scene B’s data, which no longer reflected the users’ original preferences. We argue that it is essential to promptly identify user interest shifts across scenarios. Furthermore, once such a shift occurs, it is crucial to prevent negative transfer among scenarios.

To address the aforementioned issues, we propose a new reinforcement learning-enhanced multi-scenario recommendation framework, called RUIE. Our approach leverages users’ cross-scenario behavioral sequences, enabling the model to promptly identify shifts in user interests. Based on above detection, the model could dynamically adjust sample utilization across different scene, allowing it to more effectively capture user interests in each scenes.
Moreover, extensive experiments on KuaiSAR\citep{Sun2023KuaiSAR} datasets demonstrate that RUIE outperforms existing state-of-the-art models. We further validate the effectiveness of each component in RUIE through comprehensive ablation studies.
\begin{figure}[t]
\centering
\includegraphics[width=\linewidth]{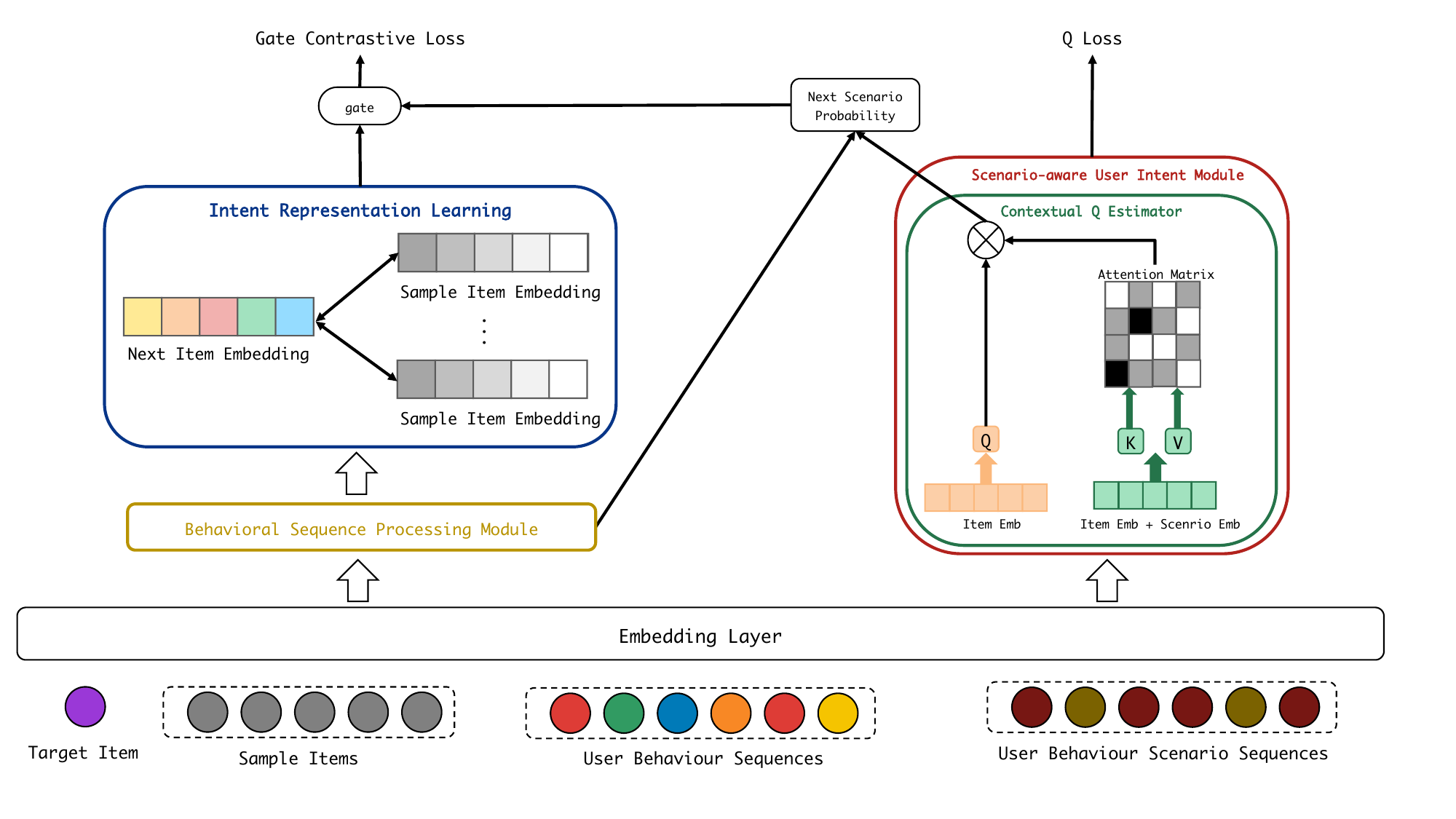}
\caption{Overall framework of our proposed RUIE.}
\label{fig:model_structure}
\end{figure}

\section{Methodology}
This section introduces the RUIE method, with the overall architecture illustrated in Figure~\ref{fig:model_structure}. To ensure that the output data exist in a unified space, we convert user sequences and scenario IDs into embeddings. To obtain the prediction of the embedding of the target item, we extract it using our behavioral sequence processing module. We also introduce Intent Representation Learning to enhance the accuracy of the embeddings.

Since user behaviors exhibit varying confidence levels across different scenarios, we leverage the Double Q-learning\citep{hasselt2010double} framework from reinforcement learning to estimate the behavior confidence (Q-value) in the scenario. We also upgrade the calculation method of the Q-value with a contextual Q Estimator, combining multi-head attention with a mask mechanism and fully connected networks. Then, we select the Q-value corresponding to the scenario of the next item (target item), which will serve as a gate to control the learning degree of the Intent Representation Learning module.

\subsection{Embedding Layer}
We utilize two embedding tables,$E_{I}$ and $E_{S}$, for all items and scenario IDs, respectively. $I$ denotes the set of all item IDs, $S$ denotes the set of all scenario IDs, and $d$ represents the embedding dimension. Given an item and a scenario, we can get the corresponding embeddings, $e_{i}$ and $e_{s}$ by lookup operation.

\subsection{Behavioral Sequence Processing Module}
In traditional recommender systems, user interest modeling is typically static. For example, historical behaviors are compressed into fixed-length vectors through simple pooling operations. However, this approach fails to capture the sequential dependencies and fine-grained patterns in user behavior. To address this issue, we adopt the NextItNet\citep{yuan2019simple} model, which employs a dilated convolutional neural network to capture long-range dependencies in user behavior sequences without information loss.

NextItNet models user behavioral sequences using a stack of residual blocks with dilated convolutions, enabling it to effectively capture both short- and long-term user interests. Unlike RNN-based or attention-based methods that may suffer from training inefficiencies or require complex alignment, NextItNet offers a simple yet effective solution by applying masked convolutions to preserve the autoregressive property. This allows the model to predict the target item conditioned on the entire history of previous behaviors.

The user behavioral sequence representation is generated as follows:

$$
\boldsymbol{\upsilon}_U(A)=f(\boldsymbol{e}_{i_1},\boldsymbol{e}_{i_2},..,\boldsymbol{e}_{i_H})
$$

where $e_1,e_2,...,e_H$ denotes the sequence of historical behavioral item embeddings with length $\mathbf{H}$. The function 
$\mathbf{f()}$ represents the stacked dilated convolutional layers with residual connections in the NextItNet architecture. The final sequence representation $\mathbf{\boldsymbol{\upsilon}_U(A)}$ captures rich contextual dependencies and is used for downstream tasks such as next-item prediction or candidate ranking.





\subsection{Scenario-aware User Intent Module}
Reinforcement learning (RL) is introduced to tackle the challenge of inconsistent user behavior across multiple scenarios. As a method where an agent learns through interaction with its environment, RL dynamically optimizes recommendation strategies based on user behaviors and feedback, better predicting the target item of user interest. Double Q-learning\citep{hasselt2010double}, a variant of Q-learning, addresses the issue of Q-learning sometimes overestimating state-action values (Q-value).

\subsubsection{Algorithm framework}
The Q-value update is shown in Algorithm~\ref{alg:doubleq}.

\begin{algorithm}
\caption{Double Q-learning}
\label{alg:doubleq}
\begin{algorithmic}[1]
\STATE Initialize \( Q^A, Q^B, s \)
\REPEAT
    \STATE Choose \( a \), based on \( Q^A(s, \cdot) \) and \( Q^B(s, \cdot) \), observe \( r, s' \)
    \STATE Choose (e.g., random) either \textsc{UPDATE}(A) or \textsc{UPDATE}(B)
    \IF{\textsc{UPDATE}(A)}
        \STATE Define \( a^* = \arg\max_a Q^A(s', a) \)
        \STATE \( Q^A(s, a) \leftarrow Q^A(s, a) + \alpha(s, a) \left( r + \gamma Q^B(s', a^*) - Q^A(s, a) \right) \)
    \ELSIF{\textsc{UPDATE}(B)}
        \STATE Define \( b^* = \arg\max_a Q^B(s', a) \)
        \STATE \( Q^B(s, a) \leftarrow Q^B(s, a) + \alpha(s, a) \left( r + \gamma Q^A(s', b^*) - Q^B(s, a) \right) \)
    \ENDIF
    \STATE \( s \leftarrow s' \)
\UNTIL{end}
\end{algorithmic}
\end{algorithm}

$Q^A,Q^B$ are the Q-value functions representing two separate estimates, this corresponds to the final Q-value in the Contextual Q Estimator. $s$ is represents the current state, which represents the user's behavior sequence feature. $a$ and $r$ denotes chosen scenario id and the reward (click,view,etc.) in the current state $s$,respectively. Q is accumulated value of r, which represents the value of long-term user engagement and will be used as the behavior confidence in the final loss. $a^*$ is the best scenario in the next state $s'$ according to the Q-value function $Q^A$. $b^*$ is the best scenario in the next state $s'$ according to the Q-value function $Q^B$.

\subsubsection{Contextual Q Estimator}
To effectively approximate the Q-value function in complex environments and enable the agent to learn more effective policies, we propose a model that integrates user behavior data with corresponding scenario identifiers. This approach aims not only to capture the migration and evolution of user interests, but also to uncover the associations between items and scenarios. By doing so, the model gains a deeper understanding of user behavior patterns and preferences across diverse contextual settings. The item embedding will add with scenario embedding, like $e_{i+s} = e_s+e_i$, it can generate richer feature representations and helping the model understand changes in scenarios. Then we will use Multi Head Attention\citep{vaswani2017attention} to model the user historical behavior. 

$$Q_i=QW_i^Q,K_i=KW_i^K,V_i=VW_i^V,i=1,\ldots,4$$

,where $Q$ is $e_i$, $K$ and $V$ is $e_{i+s}$, and $W_i^K \in \mathbb{R}^{D \times D_k}$, $W_i^V \in \mathbb{R}^{D \times D_v}$, $W_i^Q \in \mathbb{R}^{D \times D_q}$, $D_v,D_k,D_q$ is equal with $D$ divide number of attention heads $h$.

$$head_i=Attention(Q_i,K_i,V_i),i=1,\ldots,4$$

And the Attention function is $softmax(\frac{Q_iK_i^T}{\sqrt{d_k}})$.

$$MultiHead(Q,K,V)=Concat(head_1,\ldots,head_4)W^O$$

where $W^O \in \mathbb{R}^{h*D_v \times D}$, $h$ is number of attention heads.

We take the representation of the next position in the user behavior sequence as the attention representation, and then concatenate it with the behavior sequence processing results. This concatenated result will pass through two fully connected layers and a softmax activation function to obtain the final Q-value.

\subsection{Intent Representation Learning}
To enhance the coherence of items in user sequences and capture the relationship between the current behavior representation and the target item, we take the representation of the user behavior sequence module as the input to a three-layer fully connected network, and the output of the network is considered to be the prediction of the embedding of the target item. This next engaged item is considered as a positive sample. And we randomly sample 10 items (not in the sequence) as negative examples. The Euclidean distance is calculated between these pair of positive and negative examples. Then the triplet loss \citep{schroff2015facenet} is used to constrain their distances.

$$Loss(a, p, n) = max(0, \mathbf{E}(a,p)-\mathbf{E}(a,n)+\alpha)$$

where $a$ is the embedding of prediction, $p$ is the target item embedding, $n$ is negative examples embedding, $\alpha$ is temperature coefficient, and $\mathbf{E}$ is Euclidean distance function between two embeddings.

By optimizing the relative distance between samples, the distance between positive samples and the user behavior representation is minimized, prompting items in the same sequence to be more similar. Conversely, the distance between the user behavior representation and the negative samples will be required to be as large as possible and reinforce their dissimilarity.

\subsection{Loss Function}

The model predicts the target item based on the user's behavior sequence. However, learning in low-resource scenarios is often suboptimal due to imbalanced data across scenarios. To address this, we introduce a gating mechanism using the Q-value of the target item’s scenario to adjust the contrastive loss by using the Q-value of the target item's scenario. This Q-value reflects both prediction confidence and learning adequacy of target scenario. When the Q-value (denoted as next\_scenario\_prob in the formula) is low, indicating poor user interest in the scenario, the loss is down-weighted to reflect the low confidence of behavior in this scenario.

$$Gate=\frac{1}{1-\text{next\_scenario\_prob}+\epsilon}$$

This approach encourages the model to select the correct scenario, thereby reducing the loss. And $\epsilon$ is a very small positive number used to avoid the situation where the denominator is zero.

And the final loss is:

$$ Loss = Loss_q + Gate * Loss(a,p,n)$$ 

where $Loss_q$ is Double Q-learning loss, $L_(a,p,n)$ is triplet loss in Intent Representation Learning.

\section{Experiments}
In this section, offline experiments are designed to evaluate the
performance and effectiveness of RUIE.

\subsection{Dataset and Settings}
We conduct experiments on an publicly available dataset \textbf{KuaiSAR}\footnote{\url{https://kuaisar.github.io/}}, which integrates search and recommendation scenarios, encompassing authentic user interaction records gathered from the Kuaishou short-video platform, which is a prominent app in China boasting more than 400 million daily active users.


The train dataset start from 2023/05/22 and end with 2023/05/28 and test dataset focus on 2023/05/29. Meanwhile, we randomly sample 10,000 users for use. For each user, we construct a behavior sequence of length 20. Initially, since the user has no prior interactions, the sequence is padded with a special token equal to the maximum item ID plus one. The label is set to the user's first interacted item. Then, following chronological order, we apply a sliding window to update the sequence, where the label is always the target item in the interaction history.

\subsection{Evaluation Metrics}

For each user sequence, the final item is reserved for testing, while the remaining items are used for training. Instead of relying on negative sampling which is frequently criticized for introducing bias, models are assessed using a full ranking approach. We adopt top-K Normalized Discounted Cumulative Gain (NDCG@K) with K values set to {5, 10, 15, 20} as the evaluation metric.


\subsection{Implementation Details.}

In the reinforcement learning setting, we assign different rewards to four types of user behaviors: click, follow, like, and share, with corresponding rewards of 1, 3, 3, and 2, respectively. The discount factor is set to 0.5. Both the item embeddings and scenario embeddings are initialized randomly. The number of heads in the multi-head attention mechanism is set to 4. In the contrastive learning module, there are 3 fully connected network, each using the ReLU activation function. The learning rate is set to 0.01, the batch size is 256, and the number of training epochs is 50. The Adam optimizer is used for model optimization.

\subsection{Baselines}
The baselines are described as follow: \textbf{DIN}\cite{zhou2018deep}: DIN captures diverse user interests by applying attention to historical behaviors, assigning importance based on their relevance to the target item. It is deployed in many industry recommender systems. \textbf{SASRec}\citep{kang2018self}: SASRec leverages self-attention to model user behavior sequences, effectively capturing both short and long term dependencies without fixed-size history windows.\textbf{Caser}\citep{tang2018personalized}: Caser models user behavior as images, using horizontal and vertical convolutions to capture both union- and point-level sequential patterns. \textbf{GRU}\citep{chung2014empirical}: GRU-based models use gated recurrent networks to capture sequential dependencies and temporal dynamics in user behavior. \textbf{PEPNet}\citep{chang2023pepnet}: PEPNet is a multi-scenario, multi-task network that boosts personalized recommendations by integrating personalized priors. \textbf{STAR}\citep{sheng2021one}: STAR is a multi-scenario and multi-task network that uses a star topology structure to estimate different objectives across various scenarios. It is widely referenced in many academic studies.

\begin{table}[h]
\centering
\caption{Performance of different models on the Industrial dataset}
\label{performance compare}
\setlength{\tabcolsep}{3.5pt}
\begin{tabular}{lcccc}
\toprule
Method & N@5 & N@10 & N@15 & N@20 \\
\midrule
DIN & 0.0549 & 0.1372 & 0.2627 & 0.3294 \\
Caser & 0.1450 & 0.2549 & 0.4078 & 0.6313 \\
GRU & 0.4705 & 1.1176 & 1.6313 & 2.2470 \\
SASRec & 1.0980 & 2.3568 & 3.4745 & \textbf{4.8392} \\
PEPNet & 1.3568 & 2.5450 & \textbf{3.8235} & 4.0470 \\
STAR & \textbf{1.4039} & \textbf{2.7450} & 3.1568 & 4.4549 \\
\midrule
\textbf{RUIE(Ours)} & \textbf{3.8549} & \textbf{7.4901} & \textbf{11.1098} & \textbf{14.6784} \\
\bottomrule
\end{tabular}
\end{table}

\subsection{Performance Comparison}

From Table~\ref{performance compare}, we observe that RUIE significantly outperforms all competitive baseline methods. These results demonstrate that RUIE is more effective in modeling multi-scenario environments, as it can better capture user behaviors and intentions across different scenarios, leading to more accurate interest prediction.
 
Methods based on Multi-Head Attention achieve better performance than other baselines, which may be attributed to their ability to capture high-level representations of user behavior sequences in different scenarios through the attention mechanism.

Traditional sequential modeling methods (such as GRU and DIN) can capture certain sequential patterns of user behavior to some extent, but they lack the ability to model the dynamic evolution of user interests across multiple scenarios, which leads to inferior performance.

In comparison with the widely used in multi-task and multi-scenario models(PEPNet and STAR), their performance surpasses that of traditional models. However, the absence of a feedback mechanism for each item across different scenarios leads to suboptimal performance.

\begin{table}[h]
\centering
\begin{threeparttable}
\caption{Ablation study on RUIE}
\label{ablation study}
\setlength{\tabcolsep}{2.5pt}
\begin{tabular}{lcccc}
\toprule
Method & N@5 & N@10 & N@15 & N@20 \\
\midrule
\textbf{RUIE} w/o MHA\&Gate\&SUIM & 2.8588 & 5.8588 & 8.6627  & 11.7411 \\
\textbf{RUIE} w/o MHA\&Gate     & 3.2745 & 6.6196 & 9.7333  & 12.6980 \\
\textbf{RUIE} w/o MHA           & 3.5098 & 7.0039 & 10.4039 & 13.7921 \\
\textbf{RUIE}                    & 3.8549 & 7.4901 & 11.1098 & 14.6784 \\
\bottomrule
\end{tabular}
\end{threeparttable}
\end{table}

\subsection{Ablation Study}

Since the reinforcement learning module, the gating module, and the calculation of Q-value are the core of our model, we conducted the following ablation experiments to verify their effectiveness, the result is on Table~\ref{ablation study}.

 \textbf{RUIE}(w/o MHA) We disable the multi-head attention mechanism in the Q-value calculation and keep two fully connected layers in Contextual Q Estimator. The significant decline implies that Multi-Head Attention can capture the migration and evolution of user interests but also to learn the associations between different items and scenarios.
 
\textbf{RUIE}(w/o MHA\&Gate) Without gate in loss function. Reinforcement learning is not well integrated with comparative learning, and the learning direction of the model will lose an important signpost. In this case, metrics drop on the datasets.

   \textbf{RUIE}(w/o MHA\&Gate\&SUIM) This version removes Scenario-aware User Intent Module. It can be observed that Reinforcement Learning could optimizes recommendation strategies based on user behaviors and feedback, better predicting the target item of user interest.

\section{CONCLUSION}

In this work, we propose an efficient multi-scenario modeling approach to better capture user behavioral intent across diverse scenarios. The user interests evolving across multiple scenarios can be effectively captured by the Scenario-aware User Intent Module and applied to our overall multi-scenario learning. Experimental results demonstrate that RUIE outperforms several strong baselines, highlighting its potential for broader adoption in multi-scenario recommender systems. The work offers a fresh perspective on multi-scenario modeling and highlights promising directions for future research.

\bibliographystyle{ACM-Reference-Format}

\bibliography{references} 

\end{document}